\documentclass[twocolumn,aps,prl,showpacs,superscriptaddress,preprintnumbers,floatfix,nofootinbib,longbibliography]{revtex4-1}
\usepackage{multirow}
\usepackage{hhline}
\usepackage{bbm}
\usepackage{epsfig}
\usepackage{graphics}  
\usepackage{graphicx}  
\usepackage{dcolumn}   
\usepackage{bm}        
\usepackage{amssymb}   
\usepackage{amsmath}   
\usepackage{amsfonts}
\usepackage[percent]{overpic}
\usepackage{natbib}
\usepackage{hyperref}
\usepackage[usenames]{color}
\usepackage{CJK}

\newcommand{\pT} {p_{\mathrm{T}}}
\newcommand{\lr}[1]{\left\langle #1\right\rangle}

\newcommand{\Eq}[1]{Eq.~(\ref{#1})}

\usepackage{tikz,xcolor,hyperref}

\definecolor{lime}{HTML}{A6CE39}
\DeclareRobustCommand{\orcidicon}{
	\begin{tikzpicture}
	\draw[lime, fill=lime] (0,0) 
	circle [radius=0.16] 
	node[white] {{\fontfamily{qag}\selectfont \tiny ID}};
	\draw[white, fill=white] (-0.0625,0.095) 
	circle [radius=0.007];
	\end{tikzpicture}
	\hspace{-2mm}
}
\foreach \x in {A, ..., Z}{%
	\expandafter\xdef\csname orcid\x\endcsname{\noexpand\href{https://orcid.org/\csname orcidauthor\x\endcsname}{\noexpand\orcidicon}}
}
\foreach \x in {A, ..., Z}{%
	\expandafter\xdef\csname orcid\x\endcsname{\noexpand\href{https://orcid.org/\csname orcidauthor\x\endcsname}{\noexpand\orcidicon}}
}

\begin{document}

\title{Imprints of octupole collectivity in uranium-238 on relativistic heavy-ion flow observables}

\newcommand{\moe}{Key Laboratory of Nuclear Physics and Ion-beam Application (MOE), and Institute of Modern Physics, Fudan
University, Shanghai 200433, China}
\newcommand{\fudan}{Shanghai Research Center for Theoretical Nuclear Physics, NSFC and Fudan University, Shanghai 200438, China}
\newcommand{\fudanP}{Physics Department and Center for Particle Physics and Field Theory, Fudan University, Shanghai 200438, China}
\newcommand{\sbu}{Department of Chemistry, Stony Brook University, Stony Brook, NY 11794, USA}
\newcommand{\bnl}{Physics Department, Brookhaven National Laboratory, Upton, NY 11976, USA}

\author{Chunjian Zhang}\email{chunjianzhang@fudan.edu.cn}\affiliation{\moe}\affiliation{\fudan}\affiliation{\sbu}
\author{Jiangyong Jia}
\email{jiangyong.jia@stonybrook.edu}\affiliation{\sbu}\affiliation{\bnl}
\author{Jinhui Chen}
\email{chenjinhui@fudan.edu.cn}\affiliation{\moe}\affiliation{\fudan}
\author{Chun Shen}\email{chunshen@wayne.edu}\affiliation{Department of Physics and Astronomy, Wayne State University, Detroit, Michigan 48201, USA}\affiliation{RIKEN BNL Research Center, Brookhaven National Laboratory, Upton, New York 11973, USA}
\author{Lumeng Liu}
\email{liulumeng@fudan.edu.cn}\affiliation{\fudanP}

\begin{abstract}
Some atomic nuclei exhibit enhanced octupole collectivity, reflected in finite reflection-asymmetric multipole correlations rather than necessarily in a rigid static pear-shaped ground state. Low-energy studies indicate finite octupole strength in uranium-238, commonly interpreted as soft or vibrational in nature, in addition to its large prolate quadrupole collectivity~\cite{MCGOWAN1994569,KIBEDI:2002wxc}, in addition to its large prolate quadrupole collectivity. Here we investigate how such octupole correlations can be encoded in the initial geometry of relativistic heavy-ion collisions and mapped to final-state flow observables. Using state-of-the-art hydrodynamic calculations, we demonstrate quantitative sensitivity to octupole-induced features encoded in the initial-state geometry and suggest a modest octupole collectivity in uranium-238, confirmed by the latest high-energy experimental measurements~\cite{STAR:2025elk}. These findings provide as a complementary probe of odd-order nuclear collectivity and help constrain quark-gluon plasma initial conditions.
\end{abstract}

\maketitle
\label{sec:intro}

\textit{Introduction.---}
The atomic nucleus, as a strongly interacting many-body system, exhibits a variety of shapes from spherical to super-deformed~\cite{Heyde:2011pgw,Verney:2025efj}. Nuclear deformation has been traditionally inferred from low-energy spectroscopic measurements and models of reduced transition probability between low-lying rotational states, less than a few tens of MeVs/nucleon beam energies~\cite{Moller:2015fba}. More recently, it has been demonstrated that nuclear shapes can also be probed in high-energy nuclear collisions in yoctosecond-scale ($10^{-24}s$) resolution of nucleon spatial distributions by taking advantage of the strong responses of the hydrodynamics collective flow to the shape and size of the initial state~\cite{Jia:2022ozr,Jia:2025wey}. This paradigm has been validated by measurements at Relativistic Heavy Ion Collider (RHIC)~\cite{STAR:2021mii,STAR:2024wgy} and at the Large Hadron Collider (LHC)~\cite{ALICE:2021gxt,ATLAS:2022dov}, aided by significant theoretical efforts~\cite{Giacalone:2021udy,Bally:2021qys,Jia:2022qgl,Zhao:2022uhl,Ma:2022dbh,Giacalone:2023cet,Giacalone:2023hwk,Ryssens:2023fkv,Lu:2023fqd,Fortier:2024yxs,Zhao:2024lpc, Mantysaari:2024uwn, Duguet:2025hwi}.

In modeling high-energy collisions, the nuclear density is commonly parameterized by a deformed Woods-Saxon (WS) profile~\cite{Woods:1954zz,Miller:2007ri}
\begin{equation}\label{eq:1}\begin{split}
\rho(r,\theta) &\propto \left(1+\exp[(r-R(\theta,\phi))/a]\right)^{-1},\\ 
R(\theta,\phi)&=R_0\big(1+\beta_2\left[\cos\gamma Y_{2,0} + \sin\gamma Y_{2,2} \right] \\
&~~~~~+ \beta_3 Y_{3,0} + \beta_4 Y_{4,0}\big),
\end{split}
\end{equation}
where $R_0 \approx 1.2A^{1/3}$ denotes the nuclear radius, $a$ is the surface or skin depth. The nuclear surface $R(\theta,\phi)$ is expanded in spherical harmonics $Y_{l, m}(\theta, \phi)$ in the intrinsic frame, including only the most relevant axial symmetric quadrupole ($\beta_2$), octupole ($\beta_3$), and hexadecapole ($\beta_4$) deformations, while triaxiality parameter $\gamma$ ($0^{\circ} \leq \gamma \leq 60^{\circ}$) characterizes deviations from axial symmetry~\cite{bohr,Hagino:2025vxe}. In this work, the parameters $\beta_n$ in Eq.~\ref{eq:1} are used as effective amplitudes controlling multipole correlations in event-by-event coordinate-space sampling. In particular, $\beta_3$, an independent degree of freedom orthogonal to even order, should not be interpreted as evidence for a rigid static octupole deformation of the $^{238}$U ground state. Rather, it parameterizes the RMS strength of reflection-asymmetric octupole correlations that can contribute to the initial geometry sampled by an ultrarelativistic collision.

Using the ``Imaging-by-Smashing" method, the STAR experiment at RHIC has extracted effective shape parameters associated with the ground-state $^{238}$U, indicating a strongly prolate quadrupole deformation ($\beta_{2}$) and a small but finite triaxiality ($\gamma$)~\cite{STAR:2024wgy}, consistent with low-energy constraints. Model studies further suggest sensitivity to hexadecapole deformation $\beta_4$ in $^{238}$U+$^{238}$U collisions~\cite{Ryssens:2023fkv,Xu:2024bdh,Wang:2024vjf}. Here, $\beta_4$ can be correlated with the dominant $\beta_2$ in nuclear structure calculations through multipole couplings in the deformation energy surface~\cite{Moller:2015fba,Peter:1980,Lotina:2023wcy,Lotina:2024psl,Lotina:2024soz,Gupta:2023cvv,Delaroche:2009fa,Kumar:2023iki,Rodriguez-Guzman:2025feb,Inakura:2025sah}.

\begin{figure*}[htbp]
\centering
\includegraphics[width=0.6\linewidth]{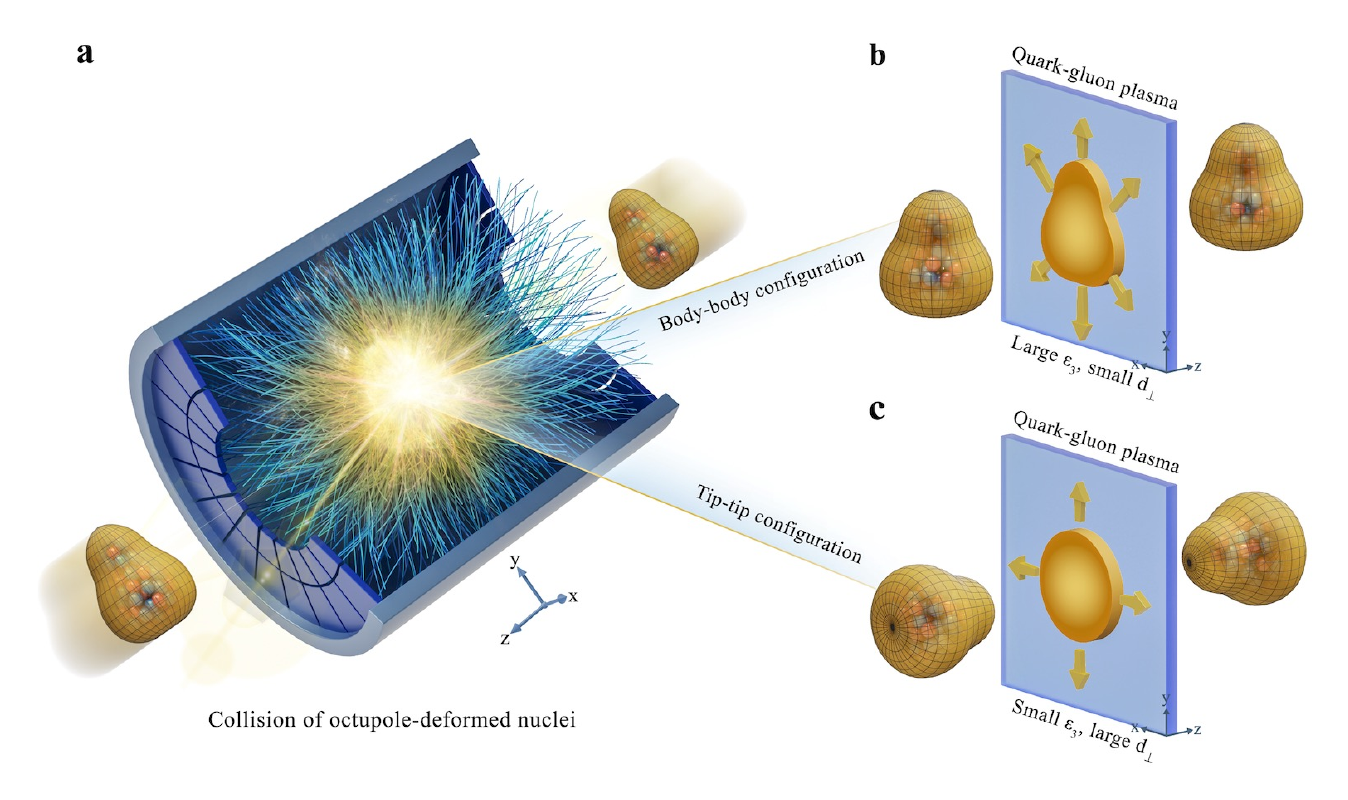}
\caption{\label{fig1}Illustration for relativistic heavy-ion collisions with sampled octupole-correlated nuclear configurations. Cartoon of collisions of involving nuclei whose event-by-event coordinate-space configurations contain reflection-asymmetric octupole correlations. The produced charged particle tracks measured in the detector (panel a). Head-on body-body (panel b) and tip-tip configurations (panel c), with corresponding 3D profile of the initially quark-gluon plasma (QGP), in which the arrows indicate the pressure gradients.}
\end{figure*}

The octupole collectivity in the heavy $^{238}$U nucleus, away from the best known region of octupole deformed nuclei centered at $Z\sim90$, $N\sim136$, remains less explored. Low-energy spectroscopic studies of $^{238}$U have long established finite octupole strength through $E3$ transition observables. However, these data do not imply a rigid static octupole-deformed ground state at zero or low spin~\cite{Spear:1989zz,Agbemava:2016mvz,Ward:1996gxf,Ahmad:1993nx,MCGOWAN1994569,KIBEDI:2002wxc,Butler:1996zz,Butler:2016rmu,Zhu:2010zzc}. The octupole collectivity in $^{238}$U is more commonly interpreted as soft or vabrational in nature~\cite{Ahmad:1993nx,Butler:1996zz,RevModPhys.91.015001,Verney:2025efj}. This octupole collectivity plays an important role in nuclear structure models~\cite{Bender:2003jk}, fission dynamics~\cite{Schunck:2015hxw,Scamps:2018vku,Huo:2024hex}, with potential implications in searches for atomic electric dipole moments~\cite{Ahmad:1993nx,Butler:1996zz,Gaffney2013,RevModPhys.91.015001,osti_2280968}. The question addressed here is different from establishing a static pear-shaped ground state. We ask whether the equal-time coordinate-space octupole correlations associated with this collectivity can leave measurable imprints on the initial geometry of relativistic heavy-ion collisions and subsequently on final-state flow observables.

\textit{Mapping Octupole correlations to triangular flow.---} The ``Imaging-by-Smashing" approach focuses on head-on (near-zero impact parameter) collisions of octupole-deformed nuclei (Fig.~\ref{fig1}a-c). Each incident nucleus is a Lorentz-contracted disc, generating a quark–gluon plasma (QGP)--a hot, dense state of deconfined quarks and gluons~\cite{STAR:2005gfr,PHENIX:2004vcz}. The shape and size of the QGP overlap region are determined by the transverse ($xy$) plane projection of the colliding systems, which directly mirror the shape of those colliding nuclei encoded in the initial-state modeling. Body–body collisions generate a larger, triangular QGP profile, whose pressure-gradient-driven asymmetric expansion is illustrated by the arrows in Fig.~\ref{fig1}b. In contrast, tip–tip collisions yield a compact, circular QGP, leading to a faster but azimuthally symmetric expansion (Fig.~\ref{fig1}c). 

The sensitivity to nuclear deformation arises from the approximately linear mapping between the initial-state geometry and final-state anisotropic flow~\cite{STAR:2024wgy}. When deformed nuclei collide in random orientations, they imprint distinct spatial anisotropies on the initial energy density profile, characterized by eccentricity coefficients $\varepsilon_{n>1} \equiv \left|\int r^n e^{i n \phi} e(r, \phi) d^2r / \int r^n e(r, \phi) d^2r \right|$, where the local energy density $e(r,\phi)$ in the $xy$ plane is derived from the overlap region, although its precise magnitude depends on the hydrodynamic model used. Driven by pressure gradients, the QGP undergoes hydrodynamic collective expansion, transferring the $\varepsilon_n$ into azimuthal anisotropy of final-state hadrons, described by a Fourier expansion of the particle azimuthal angle distribution $d N / d \phi \propto 1+2 \sum v_n \cos n\phi$~\cite{Poskanzer:1998yz,Borghini:2000sa,Heinz:2013th,Chen:2024aom}. The elliptic ($v_2$) and triangular ($v_3$) flows linearly respond to the initial eccentricities $\varepsilon_n$~\cite{Niemi:2015qia}, $v_n \propto \varepsilon_n$, making them sensitive to quadrupole and octupole collectivity.

Event-wise average transverse momentum, ($\langle p_{\rm T}\rangle$), quantifies QGP compactness via the inverse area of the overlap $d_{\perp} \propto 1 / \sqrt{\left\langle x^2\right\rangle\left\langle y^2\right\rangle}$~\cite{Schenke:2020uqq} and drives radial flow, following a linear relation, $\delta p_T \propto \delta d_{\perp}$, where $\delta p_{\mathrm{T}}=\langle p_{\mathrm{T}}\rangle-\langle\!\langle \pT\rangle\!\rangle$ and $\delta d_{\perp}=d_{\perp}-\left\langle d_{\perp}\right\rangle$ denote event-wise deviations from mean values~\cite{Bozek:2012fw,STAR:2024wgy}. Here, $\langle\!\langle \pT\rangle\!\rangle$ represents the event averaged transverse momentum in event ensemble. 

In collisions of spherical nuclei, $\varepsilon_2$ primarily reflects the impact-parameter-driven overlap ellipticity, while non-zero $\varepsilon_3$ arises solely from random fluctuations of nucleons~\cite{Alver:2010gr,Teaney:2010vd,Bozek:2013uha}, leading to an anti-correlation $\lr{v_3^2} \sim 1/A$ in the absence of octupole deformation ($\beta_3=0$). The presence of $\beta_3$ and $\beta_4$ enhances the variance of $v_n$, following a quadratic form~\cite{Jia:2021tzt} for $n=2,3$, 
\begin{align}\nonumber
\lr{v_2^2}&\approx a_2+ b_{2,2} \beta_2^2+b_{2,3} \beta_3^2+b_{2,4} \beta_2\beta_4,\\\label{eq:2}
\lr{v_3^2}&\approx a_3+ b_{3,3} \beta_3^2+b_{3,4} \beta_4^2.
\end{align}
$\beta_4$ can affect $v_2$ by its non-linear coupling with $\beta_2$~\cite{Ryssens:2023fkv}, while $v_3$ is not expected to be affected by $\beta_2$~\cite{Jia:2021tzt}. Here, $a_n$ encodes the system size and centrality dependence for spherical nuclei, while the coefficients $b_{n , m}$ are expected to be $A$ independent. These relations enable a data-driven method to constrain the $\beta_n$ parameters by comparing collisions of two species of similar sizes; see Refs.~\cite{Giacalone:2021udy,Jia:2021tzt} for details. 

In this Letter, we identify experimental signatures sensitive to octupole collectivity in $^{238}$U using state-of-the-art hydrodynamic calculations. We provide quantitative constraints on the magnitude of $\beta_{3,\rm U}$ through two novel observables, triangular flow and its correlation with transverse momentum, offering new insights into pear-shaped collectivity that can be contrasted with recent high-precision experimental measurements~\cite{STAR:2025elk,STAR:2002eio}.

We emphasize that for a rotational invariant Hamiltonian, the one-body density of a $J^\pi=0^+$ ground state is spherical in the laboratory frame~\cite{Dobaczewski:2025rdi}. While intrinsic deformations are not directly observable, collective multipole correlations in even-even nucleus can be inferred from electric transition rates ($B(E\ell)$), within theoretical models~\cite{Ryssens:2023fkv}. In our approach, the deformation parameters $\beta_n$ serve as effective intrinsic-shape descriptors in event-by-event hydrodynamic modeling, characterizing the strength of multipole correlations rather than fixed laboratory-frame orientation or static symmetry breaking~\cite{Woods:1954zz,Miller:2007ri}. The observables studied here are multi-particle correlation measures, sensitive to event-by-event fluctuations and deformation collectivity encoded in the collision's initial state, for which orientation-interference effects decay exponentially with increasing mass number~\cite{Ke:2025tyv}.

\textit{Hydrodynamic model and observables.---}\label{subsec:hydro}
To identify experimental signatures of octupole deformation in $^{238}$U, we simulate event-by-event (2+1)D boost-invariant $^{238}$U+$^{238}$U collisions at $\sqrt{s_{\rm NN}}=$ 193 GeV and $^{197}$Au+$^{197}$Au collisions at $\sqrt{s_{\rm NN}}=$ 200 GeV employing the IP-Glasma+MUSIC+UrQMD framework. This model consists of an initial state IP-Glasma model~\cite{Schenke:2012wb,Mantysaari:2017cni}, coupled with the hybrid framework of relativistic viscous hydrodynamics, MUSIC~\cite{Schenke:2010nt,Schenke:2010rr,Paquet:2015lta}, followed by hadronic transport, UrQMD~\cite{Bass:1998ca,Bleicher:1999xi}. The energy density $e(r,\phi)$ from the Yang-Mills evolution is used to initialize the hydrodynamical evolution. This model has been tuned to successfully reproduce collective flow in both small and large systems at RHIC and the LHC~\cite{Schenke:2020mbo}. 

For $^{197}$Au, an odd-$A$ nucleus, we adopt $\beta_{2,\rm{Au}} = 0.14$ and triaxiality $\gamma_{\rm{Au}} = 45^\circ$, guided by conservative low-energy estimates and model systematic comparisons with neighboring nuclei~\cite{Bally:2023dxi,Wu:1996zzb}. While its odd-$A$ character introduces polarization effects associated with the unpaired nucleon that complicate spectroscopic extraction of deformation parameters~\cite{Werner:1994ojv}, the bulk geometry of heavy nuclei is dominated by collective mean-field properties~\cite{Moller:2015fba,Bally:2014jsa,Bally:2023dxi}.

To isolate the response to octupole correlations in $^{238}$U, we vary an effective octupole amplitude, denoted $\beta_{3,\rm{U}}$ from $0$ to $0.2$ while fixing $\beta_{2,\rm{U}} = 0.28$ and $\gamma_{\rm{U}} = 0^\circ$ in Eq.~\ref{eq:1} (Tab.~\ref{tab:1}). Hexadecapole effects are systematically tested by scanning $\beta_{4,\rm{U}}$ between $0$ and $0.09$~\cite{Bemis:1973zza,Zumbro:1984zz,Moller:2015fba}. The minimum nucleon distance in nuclei $d_{min}$ is 0.9 fm in these two collision systems. For each configuration, $100$k--$400$k high-statistic hydrodynamic events are generated, allowing the precise determination of shape differences between these two species. Each event is oversampled at least $100$ times during the UrQMD stage to ensure enough resolution for anisotropic flow vectors and minimize statistical fluctuations in the hadronic transport. 

$\beta_n$ parameterizes the intrinsic-shape component of the event-by-event initial geometry in deformed WS sampling. Mapping it to laboratory-frame spectroscopic observables requires additional nuclear-structure input, such as symmetry restoration or rotational model relations~\cite{Ryssens:2023fkv}, which is beyond the scope of this Letter.

We study two-particle triangular flow, $\lr{v_3^2}$, and its correlation with transverse momentum, $\left\langle v_3^2 \delta p_{\mathrm{T}}\right\rangle$, defined as
\begin{equation}\label{eq:3}\begin{split}
\langle v_3^2 \rangle &= \langle \cos [3(\phi_i-\phi_j)] \rangle_{i,j}, \\
\langle v_3^2 \delta \pT \rangle &= \langle \cos[3(\phi_i-\phi_j)] (p_{\rm{T},k} - \langle\!\langle \pT\rangle\!\rangle)\rangle_{i,j,k}.
\end{split}
\end{equation}
Here, $\phi_i$ and $p_{\mathrm{T},i}$ denote the azimuthal angle and transverse momentum of particle $i$. The particle indices $i$, $j$, and $k$ are chosen such that only unique combinations are used. The averages are first over unique particle multiplets in an event, then over all events in a fixed centrality class. $\langle v_{3}^2\rangle$ is obtained using the two-subevent method with pseudorapidity gap $|\Delta\eta|>$ 1 and the covariance $\langle v_3^2 \delta \pT \rangle$ is calculated via standard method. Results are verified across four different $\pT$ intervals.

\begin{table}[t]
\centering
\caption{\label{tab:1} Woods-Saxon parameters used in IP-Glasma+MUSIC+UrQMD simulations of $^{238}$U+$^{238}$U and $^{197}$Au+$^{197}$Au collisions. $d_{\mathrm {min}}$ represents the nucleon minimum distance.\\}
	\renewcommand{\arraystretch}{1.6}
	\setlength{\tabcolsep}{0.9mm}
\hspace*{-0.3cm}\footnotesize{
\begin{tabular}{c|c|c|c|c|c|c|c}
\hline  
Species                   & R(fm)              & a (fm)      & $d_{\mathrm {min}}$(fm) & $\beta_2$      &   $\beta_3$  & $\beta_4$      & $\gamma$ ($^{\circ}$) \\\hline
$^{197}$Au                 & 6.62   & 0.52              & 0.9                  & 0.14  &0 &  0       &  45 \\\cline{1-8}
\multirow{3}{*}{$^{238}$U} & \multirow{3}{*}{{6.81}} & \multirow{3}{*}{{0.55}} &\multirow{3}{*}{{0.9}} & \multirow{3}{*}{{0.28}} & 0.00, 0.05           & \multirow{3}{*}{0.09, 0}   & \multirow{3}{*}{{0}} \\
                          &                                  &            &               &           & 0.10, 0.15       &               \\
                          &                                  &            &                   &       & 0.20  &               &   \\
\hline
\end{tabular}}\normalsize
\end{table}

\begin{figure*}[htbp]
\centering
\includegraphics[width=0.35\linewidth]{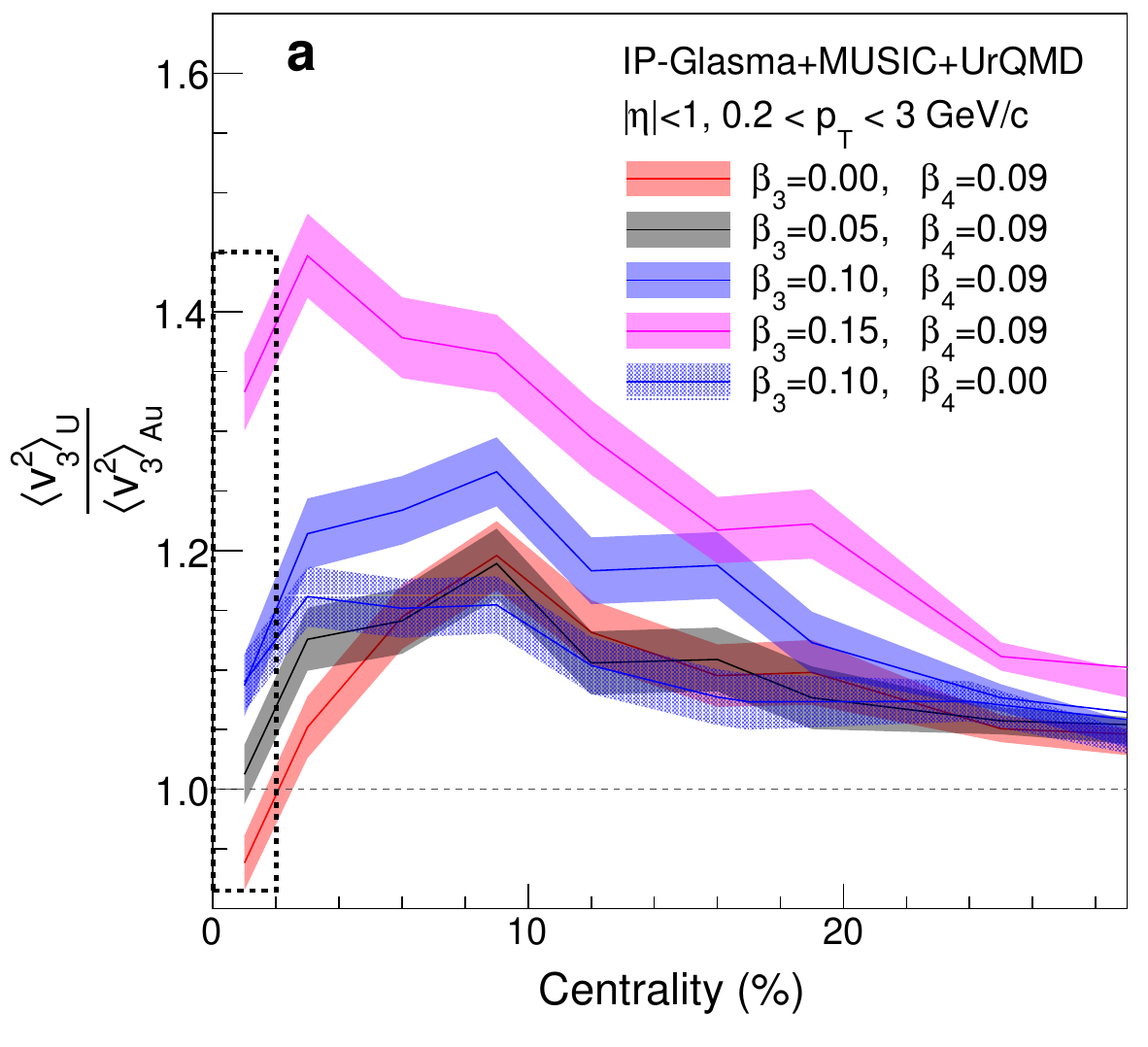}
\includegraphics[width=0.35\linewidth]{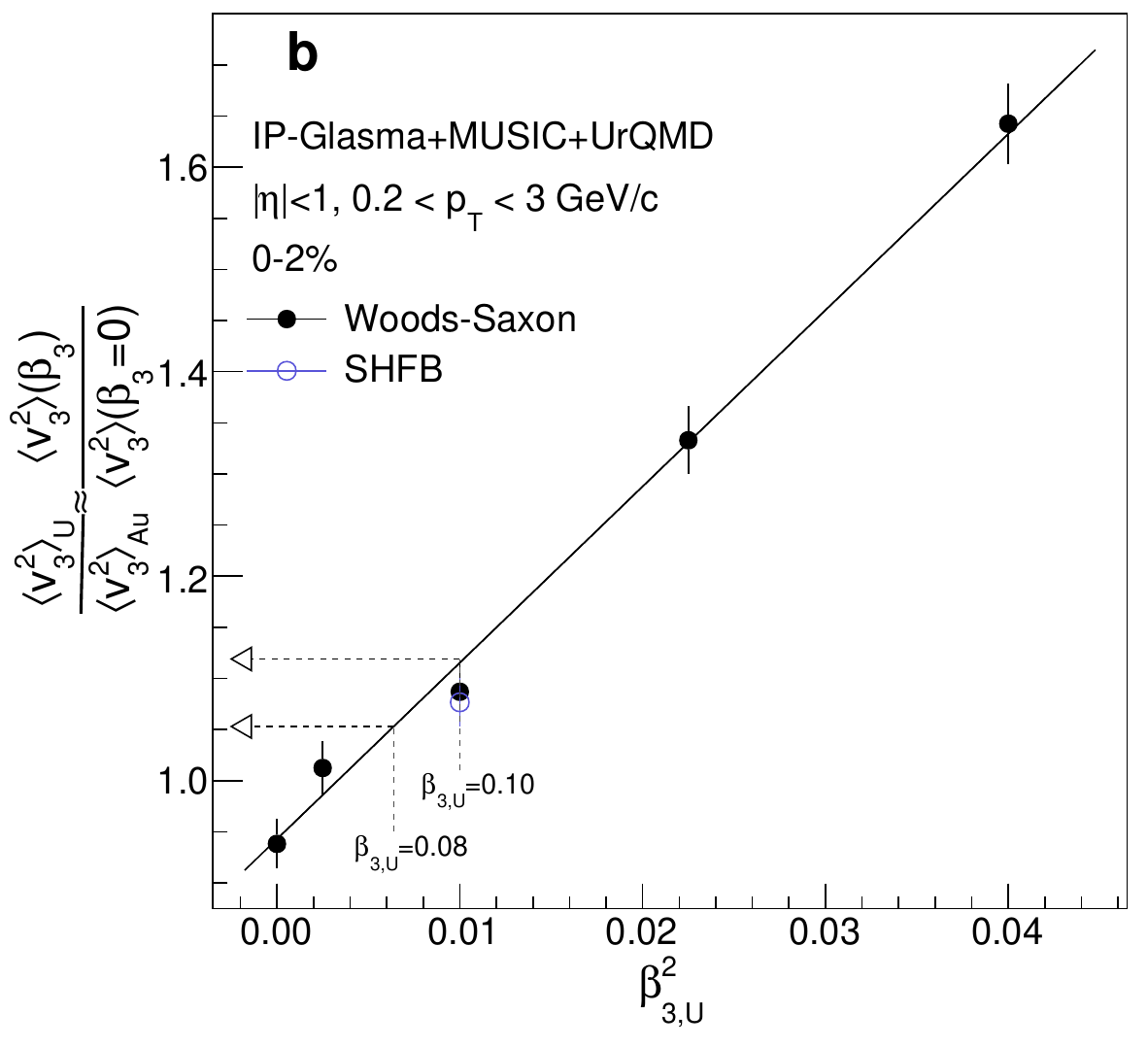}
\caption{\label{fig2}a, Ratios of $\lr{v_3^2}$ between $^{238}$U+$^{238}$U and $^{197}$Au+$^{197}$Au collisions as a function of centrality for different $\beta_{3}$ from IP-Glasma+MUSIC+UrQMD using WS densities. Shaded bands denote statistical uncertainties. The dashed box marked the 0$–$2\% centrality region that is most sensitive to $\beta_3$. b, The ratio $\lr{v_{3}^2}_{\rm{U}}/\lr{v_{3}^2}_{\rm{Au}}$ versus $\beta_{3,\rm{U}}^2$ in 0$-$2\% centrality, assuming $\beta_{3,\rm{Au}}^2=$ 0. Arrows mark $\beta_{3,\rm{U}}=$ 0.08 and 0.10 from low-energy studies~\cite{Agbemava:2016mvz,MCGOWAN1994569}, in comparable with recent STAR measurements~\cite{STAR:2025elk}. Results with SHFB density yield consistent trends.}
\end{figure*}

The observables in \Eq{eq:3} are calculated in isobar-like $^{238}$U+$^{238}$U and $^{197}$Au+$^{197}$Au collisions with near $A$, from which we construct the ratios
\begin{align}\label{eq:4}
    &R_{v_3^2}  = \frac{\lr{v_3^2}_{\rm U+U}}{\lr{v_3^2}_{\rm Au+Au}} \approx \frac{a_{3\rm U}}{a_\text{3Au}}+\frac{b_{3,3}}{a_\text{3Au}}\beta_3^2+\frac{b_{3,4}}{a_\text{3Au}}\beta_4^2,\\\label{eq:5}
    &R_{v_3^2\delta \pT} = \frac{\lr{v_3^2\delta \pT}_{\rm U+U}}{\lr{v_3^2\delta \pT}_{\rm Au+Au}} \approx a-b\beta_2\beta_3^2.
\end{align}
These ratios suppress most system-dependent effects, including final-state and odd-$A$ specific effects, sensitive to the RMS strength of octupole correlations rather than to the sign or fixed orientation of a static octupole moment. Their right-hand sides, linked to two- and three-body nucleon distributions in the intrinsic frame from \Eq{eq:1}, have a simple parametric dependence on shape parameters~\cite{Jia:2021tzt,Wang:2024ulq}, with $a$ and $b$ in Eq.~(\ref{eq:5}) being empirical positive coefficients varying with centrality~\cite{Wang:2024ulq}. The sensitivity of $R_{v_3^2\delta \pT}$ to octupole correlations requires the presence of large quadrupole component, as in $^{238}$U, making it a suitable system to probe $\beta_3$. This particular sensitivity would not appear in $^{96}$Zr+$^{96}$Zr vs $^{96}$Ru+$^{96}$Ru comparison to constrain the $\beta_{3,\rm Zr}$, given the small values of $\beta_{2,\rm Zr}$.

In Eqs.~(\ref{eq:4}) and (\ref{eq:5}), higher-order deformations in $^{197}$Au are assumed negligible based on low-energy studies. Otherwise, the constraints derived should be treated as the difference between $^{238}$U and $^{197}$Au, e.g., replacing $\beta_3^2$ by $\beta_{3,\rm U}^2$-$\beta_{3,\rm Au}^2$.

In the absence of deformation, the smaller $^{197}$Au+$^{197}$Au system are expected to exhibit larger fluctuations than $^{238}$U+$^{238}$U, yielding $R_{v_3^2} \approx a_{3,\rm U}/a_{\rm3,Au}< 1$, as well as $R_{v_3^2\delta \pT} \approx a<1$. The presence of $\beta_3$ then is to increase the values of $R_{v_3^2}$ in central collisions to exceed unity--an inversion of the hierarchy that provides a unique signature for $\beta_3$--and reduces $R_{v_3^2\delta \pT}$ further below unity, eventually inducing anti-correlation behavior at large $\beta_3$ magnitudes. Meanwhile, $\beta_4$ is expected to increase $R_{v_3^2}$ in non-central collisions~\cite{Jia:2021tzt}, while its role in $R_{v_3^2\delta \pT}$ is unclear. Our findings are discussed in light of these mechanisms in mind.


\textit{Results and discussions.---}\label{sec:results} Three distinct features are identified: 1) For $\beta_{3,\rm{U}}\gtrsim 0.05$, the ordering of $\lr{v_3^2}$ in ultra-central collisions (UCC, 0--2\% centrality) is reversed between $^{238}$U+$^{238}$U and $^{197}$Au+$^{197}$Au, i.e., $R_{v_3^2}>1$. 2) In the 2--20\% centrality range, $\lr{v_3^2}$ is still strongly enhanced by $\beta_3$. 3) $\beta_4$ increases the values of $\lr{v_3^2}$ in 2--15\% centrality, but has no impact in 0--2\% centrality, consistent with findings from the initial Glauber model~\cite{Jia:2021tzt}. It should be noted that a similar behavior has also been observed in a multi-phase transport model calculations as cross-checks~\cite{Zhang:2021kxj}. These trends follow the expectation of \Eq{eq:4} well, making them robust features that can be verified experimentally. Such a hierarchical reversal feature in 0--2\% centrality can be used to set a reliable limit on $\beta_3$.

The result of Fig.~\ref{fig2}a demonstrates that finite octupole-correlation strength in $^{238}$U, enhances the event-by-event initial-state triangularity ($\varepsilon_3$) beyond the fluctuation-driven baseline. The statistical uncertainties of current simulations, represented by the shaded bands, are small enough to be sensitive to a modest $\beta_{3,\mathrm{U}}$, when compared to experimental measurements~\cite{STAR:2024wgy}.

\begin{figure*}[htbp]
\centering
\includegraphics[width=0.35\linewidth]{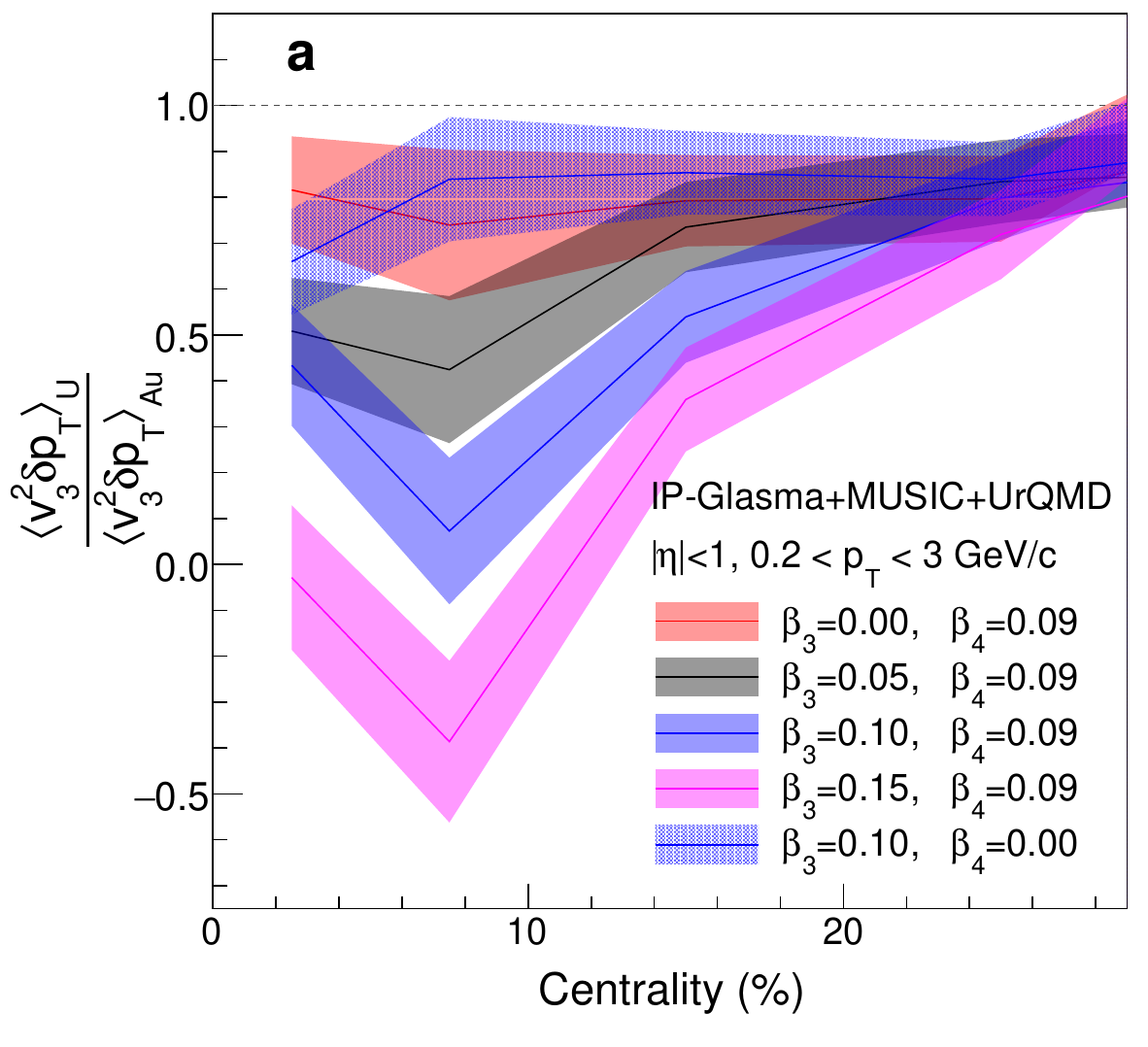}
\includegraphics[width=0.35\linewidth]{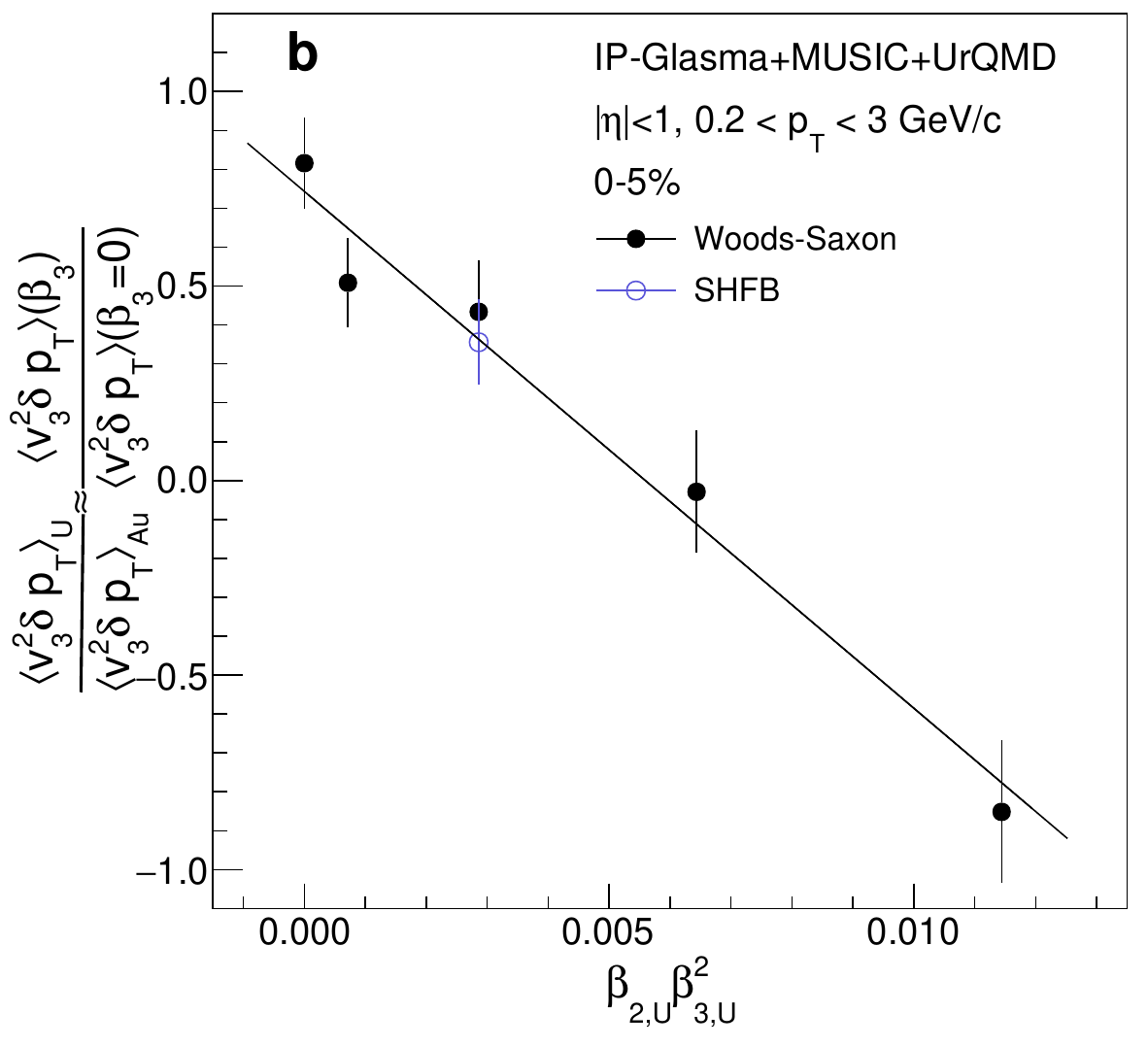}
\caption{\label{fig3}a, Ratios of $\langle v_3^2\delta\pT\rangle$ between $^{238}$U+$^{238}$U and $^{197}$Au+$^{197}$Au collisions versus centrality for different $\beta_{3,\rm{U}}$ from IP-Glasma+MUSIC+UrQMD calculations. Shaded bands denote the statistical uncertainties. b, Ratio $\lr{v_{3}^2\delta\pT}_{\rm{U}}/\lr{v_{3}^2\delta \pT}_{\rm{Au}}$ versus $\beta_{2, \rm{U}}\beta_{3,\rm{U}}^2$ for 0$-$5\% centrality. Results with SHFB density yield consistent trends.
}
\end{figure*}

Next, we focus on the UCC 0--2\% centrality range, where $R_{v_3^2}$ is mainly sensitive to $\beta_3$. As shown in Fig.~\ref{fig2}b, the predicted ratios exhibit linear dependence on $\beta_{3,\rm{U}}^2$. Assuming the magnitude of octupole strength $\beta_{3,\rm{U}}=$ 0.08 -- 0.10~\cite{Agbemava:2016mvz,MCGOWAN1994569,KIBEDI:2002wxc}, deduced from low-energy nuclear structure, we predict $R_{v_3^2} \approx$ 1.05 -- 1.12 indicated by the arrows. This range is consistent with the current STAR measurement~\cite{STAR:2025elk}. Note that experimental results falling within this range provide a strong high-energy signature of octupole collectivity in $^{238}$U. Future measurements sub-percent-level precision, combined with additional sensitivity analyses could further tighten constraints on $\beta_3$.

As a robustness test of the hydrodynamic response, We also perform constrained Skyrme Hartree-Fock-Bogoliubov (SHFB) calculations using the HFBTHO~\cite{Stoitsov:2012ri}. These calculations are not used to establish a static octupole-deformed ground state of $^{238}$U, nor do they describe octupole vibrations by themselves. Mean-field predictions of octupole minima in actinides, including $^{238}$U, are known to be functional and model-dependent~\cite{Cao:2020rgr}. Instead, for prescribed multipole moments, the constrained SHFB densities provide self-consistent radial density profiles that can be used as alternative inputs to the IP-Glasma+MUSIC+UrQMD framework. The agreement between the deformed WS and constrained SHFB inputs demonstrates that the predicted flow response is not an artifact of the WS radial parameterization. The intrinsic axial multipole moments are extracted from the SHFB density $\rho_{\rm WS}(\vec r)$ as $Q_\lambda=\int \rho_{\rm WS}(\vec r)\, r^\lambda Y_{\lambda0}(\theta)\, d^3r$, from which the deformation parameters $\beta_\lambda^{\rm DFT}=4\pi Q_\lambda/(3N_\tau R^\lambda)$ are defined. The SLy4 interaction~\cite{Chabanat:1997un} is employed in the multi-constraint SHFB calculations. We find that WS and realistic SHFB densities lead to consistent results, indicating that WS parameterization is sufficient to capture the deformation features relevant for the $^{238}$U nucleus, as shown in Fig.~\ref{fig2}b. 

To gain further insight into nuclear shape, we examine  the three-particle correlator $\langle v_3^2 \delta\pT \rangle$. Body–body collisions create a small $d_{\perp}$ with large $\varepsilon_3$ QGP shape, while tip-tip collisions form a large $d_{\perp}$ with minimal $\varepsilon_3$ profile, as illustrated in Fig.~\ref{fig1}b and c, leading to a pronounced anti-correlation between $d_{\perp}$ and $\varepsilon_3$ and resulting in different pressure-gradient-driven expansion patterns. Figure~\ref{fig3}a shows $R_{v_3^2\delta \pT}$ versus centrality for different $\beta_{3,\rm{U}}$. In central collisions (0-5\% centrality), $R_{v_3^2 \delta p_{\mathrm{T}}}$ exhibits a near-linear dependence on $\beta_{2,\mathrm{U}} \beta_{3,\mathrm{U}}^2$, aligning with the parametric form predicted by \Eq{eq:5}, and provides complementary insights to $\beta_{3,\rm{U}}$ constraints. The observed suppression is qualitatively consistent with the expected negative contribution of $\beta_{3,\rm{U}}$ to $\langle v_3^2\delta\pT\rangle$.

It is observed that $R_{v_3^2\delta \pT}$ also exhibits sensitivity to $\beta_{4,\mathrm{U}}$. A pure $\beta_{4,\mathrm{U}}=0.09$ has a negative contribution comparable to that from $\beta_{3,\mathrm{U}}=0.10$ alone. However, when both $\beta_{4,\mathrm{U}}$ and $\beta_{3,\mathrm{U}}$ are present, their combined impact is larger. This suggests that terms like -$\beta_2\beta_3\beta_4$ and -$\beta_2\beta_4^2$ should be added to \Eq{eq:5} in non-central collisions. These couplings imply that the signature of $\beta_{3,\rm{U}}$ and $\beta_{4,\rm{U}}$ requires larger $\beta_{2,\rm{U}}$, making $^{238}$U an ideal nucleus. Moderate values of $\beta_{3,\mathrm{U}}$ and $\beta_{4,\mathrm{U}}$ ($\beta_{3,\mathrm{U}} \lesssim 0.1$, $\beta_{4,\mathrm{U}}\approx$ 0.05 -- 0.1), $R_{v_3^2\delta \pT}$ falls in the range of 0.5--1 in central collisions, underscoring the need for combined observables, such as $R_{v_3^2}$ and $R_{v_3^2\delta \pT}$, to simultaneously disentangle $\beta_3$ and $\beta_4$. 

We validate the linear scaling of $R_{v_3^2\delta\pT}$ in 0--5\% centrality, shown in Fig.~\ref{fig3}b as a function of $\beta_{2,\rm{U}}\beta_{3,\rm{U}}^2$. This linear scaling serves as another significant constraint on octupole deformation in $^{238}$U in high-energy collisions, compared to the recent STAR measurements~\cite{STAR:2025elk}. Importantly, $R_{v_3^2\delta \pT}$ also exhibits consistent agreement between WS and SHFB density input.
 \begin{figure*}[htbp]
\centering
\includegraphics[width=0.7\linewidth]{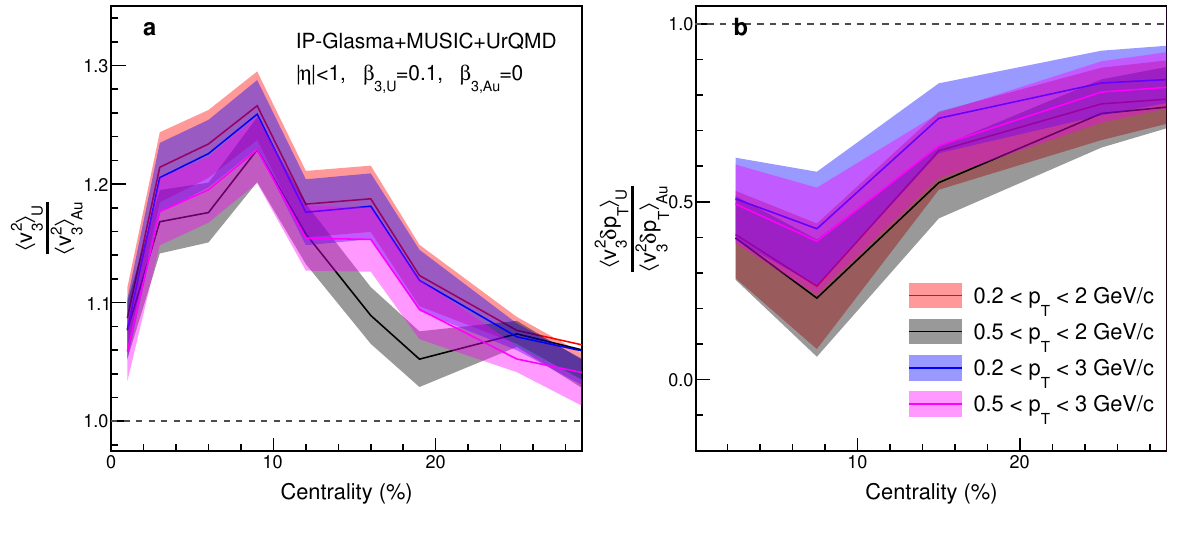}
\caption{\label{fig4}Centrality dependence of ratios of $\lr{v_3^2}$ (panel a) and $\langle v_3^2\delta\pT\rangle$ (panel b) between $^{238}$U+$^{238}$U and $^{197}$Au+$^{197}$Au collisions from the IP-Glasma+MUSIC+UrQMD model for four $\pT$ intervals. The shaded bands denote the statistical uncertainties.}
 \end{figure*}
 
 Figure~\ref{fig4} shows the centrality dependence of $R_{v_3^2}$ (panel a) and $R_{v_3^2\delta\pT}$ (panel b) calculated in four $\pT$ intervals with fixed $\beta_{3,\rm{U}}=0.10$. No variations with the choice of $\pT$ interval are observed. This implies that the final-state effects are almost completely canceled, leaving the model uncertainties mainly due to the initial conditions. The 20\% difference in the mass numbers between $^{197}$Au and $^{238}$U does not lead to any appreciable difference in the hydrodynamic response. The stability of the ratio against varying $\pT$ intervals is crucial to demonstrate that the impact of nuclear deformation can be measured without the need to impose stringent $\pT$ selections.
 
\textit{Summary.---} We have investigated how effective octupole correlations in $^{238}$U can be encoded in the initial geometry of relativistic heavy-ion collisions and manifested in final-state flow observables. The parameter $\beta_3$ used here represents an effective RMS octupole-correlation amplitude in event-by-event initial-state modeling, rather than a rigid static octupole deformation of the $^{238}$U ground state. By constructing ratios of the same observable between collisions of $^{238}$U+$^{238}$U and $^{197}$Au+$^{197}$Au within IP-Glasma+MUSIC+UrQMD framework, we identify quantitative signatures associated with octupole collectivity characterized by $\beta_{3,\rm{U}}$. In particular, a modest $\beta_{3,\rm{U}}$ leads to a characteristic reverse ordering of $\lr{v_3^2}$ ratio in UCC 0--2\% centrality, while hexadecapole collectivity parameterized by $\beta_{4,\rm{U}}$ primarily affects non-central collisions. The consistency with recent STAR measurements provides high-energy support for a finite effective octupole-correlation strength in $\beta_{3,\rm{U}}\sim 0.08-0.1$~\cite{STAR:2025elk}. We further show that the ratio of $\lr{v_3^2\delta \pT}$ exhibits complementary sensitivity to both $\beta_{3,\rm{U}}$ and $\beta_{4,\rm{U}}$ in a centrality-dependent manner. $\langle v_3^2\rangle$ increases linearly with $\beta_3^2$, while $\langle v_3^2 \delta\pT \rangle$ exhibits a previously unexplored suppression, scaling linearly with $\beta_2\beta_3^2$. Importantly, we confirm that WS and SHFB densities yield consistent results in heavy nuclei. 

Taken together, the combined analysis of these ratios enables systematic constraints on higher-order collectivity, including an upper limit on $\beta_{4,\rm{U}}$, once $\beta_{3,\rm{U}}$ is determined. Their ratios are insensitive to $p_{\rm T}$ cuts, demonstrating robustness against kinematic selections. Overall, these observables can be combined with Bayesian analysis to further refine quantitative constraints. 

By providing new constraints on initial-state modeling, this ``Imaging-by-Smashing" approach contributes to reducing uncertainties in the description of QGP initial conditions, and their subsequent dynamics and transport properties~\cite{PhysRevC.103.054904,Nijs:2020ors,Jia:2022ozr}. Moreover, these findings provide a complementary probe of non-zero octupole collectivity in heavy nuclei~\cite{Butler:1996zz,Butler:2016rmu}, complementing traditional nuclear-structure studies. While low-energy measurements indicate finite octupole collectivity in $^{238}$U~\cite{MCGOWAN1994569,KIBEDI:2002wxc}, the interpretation in terms of static rigid or soft vibrational modes remains model dependent and under debate~\cite{Rodriguez-Guzman:2020lky,Robledo:2012du,PhysRevC.88.051302}. Future high-energy measurements of multi-particle flow fluctuations, such as $(v_3\{4\}/v_3\{2\})^4$, as explored in Refs.~\cite{Liu:2025fnq,Dimri:2023wup}, provide complementary and experimentally accessible probe of the sensitivity to these scenarios.

\textit {Acknowledgements.---} We thank P. A. Butler, G. Giacalone, P. Garrett, and
K. Hagino for valuable comments and discussions. This work is supported in part by the National Key Research and Development Program of China under Contract Nos. 2024YFA1612600, 2022YFA1604900, the National Natural Science Foundation of China (NSFC) under Contract Nos. 12025501, 12547102, 12205051, the Natural Science Foundation of Shanghai under Contract No. 23JC1400200, Shanghai Pujiang Talents Program under Contract No. 24PJA009, China Postdoctoral Science Foundation under Grant No. 2024M750489.
J. Jia and C. Shen are supported by the U.S. Department of Energy, Office of Science, Office of Nuclear Physics, under DOE Awards No. DE-SC0024602 and DE-SC0021969, respectively. C. Shen acknowledges a DOE Office of Science Early Career Award.
\bibliography{ref}

\end{document}